
\input epsf
%
\input harvmac
\def\Title#1#2{\rightline{#1}\ifx\answ\bigans\nopagenumbers\pageno0\vskip1in
\def\acknowledgments{\bigskip\centerline{\bf Acknowledgments}\smallskip}
\else\pageno1\vskip.8in\fi \centerline{\titlefont #2}\vskip .5in}

scaled\magstep3
 
scaled\magstep3
%
%
\ifx\epsfbox\UnDeFiNeD\message{(NO epsf.tex, FIGURES WILL BE IGNORED)}
\def\figin#1{\vskip2in}
\else\message{(FIGURES WILL BE INCLUDED)}\def\figin#1{#1}\fi
\def\ifig#1#2#3{\xdef#1{fig.~\the\figno}
\goodbreak\midinsert\figin{\centerline{#3}}%
\smallskip\centerline{\vbox{\baselineskip12pt
\advance\hsize by -1truein\noindent\footnotefont{\bf Fig.~\the\figno:}
#2}}
\bigskip\endinsert\global\advance\figno by1}

\def\ifigure#1#2#3#4{
\midinsert
\vbox to #4truein{\ifx\figflag\figI
\vfil\centerline{\epsfysize=#4truein\epsfbox{#3}}\fi}
\narrower\narrower\noindent{\footnotefont
{\bf #1:}  #2\par}
\endinsert
}
%
\def\frac#1#2{{#1 \over #2}}
\font\ticp=cmcsc10
%
%
\gdef\journal#1, #2, #3, 1#4#5#6{{\sl #1} {\bf #2}
(1#4#5#6) #3}
\lref\lpst{D. A. Lowe, J. Polchinski, A. Strominger and L. Thorlacius,
unpublished (1994).}
\lref\aslh{A. Strominger, Les Houches Lectures on Black Holes, to appear.}
\lref\caru{V.A.~Rubakov, \journal Nucl.~Phys., B203, 311, 1982;
C.G.~Callan,  \journal Phys.~Rev., D25, 2141, 1982; \journal ibid, D26,
2058, 1982; \journal Nucl.~Phys., B212, 391, 1983.}
\lref\bps{T.~Banks, M.~Peskin, and L.~Susskind, \journal
Nucl.~Phys., B244, 125, 1984.}
\lref\jpia{ J.~Polchinski,
 \journal Nucl.~Phys., B242, 345, 1984; I.~Affleck and J.~Sagi,
\journal Nucl.~Phys., B417, 374, 1994.}
\lref\polst{J.~Polchinski and A.~Strominger, {\sl A Possible
Resolution of the Black Hole Information Puzzle}, Preprint UCSBTH-94-20,
hep-th/9407008.}
\lref\hawk{S.W.~Hawking, \journal Comm.~Math.~Phys., 43, 199, 1975.}
\lref\stas{A.~Strominger and S.P.~Trivedi, \journal Phys. Rev., D48,
5778, 1993.}
\lref\dxbh{S.B.~Giddings and A.~Strominger, \journal Phys. Rev., D46,
627, 1992.}
\lref\bos{T.~Banks, M.O'Loughlin, and A.~Strominger, \journal
Phys.~Rev., D47, 4476, 1993; see also T.~Banks and M.O'Loughlin, \journal
Phys.~Rev., D47, 540, 1993.}
\lref\garst{D.~Garfinkle and A.~Strominger, \journal Phys.~Lett., B256,
146, 1991.}
\lref\sbg{S.B.~Giddings, \journal Phys.~Rev., D49, 4078, 1994;
\journal Phys.~Rev., D49, 947, 1994.}
\lref\remn{F.~Dowker, J.P.~Gauntlett,
D. Kastor and J. Traschen,\journal Phys.~Rev., D49, 2909, 1994;
F.~Dowker, J.P.~Gauntlett, S.B.~Giddings, and G.~Horowitz,
\journal Phys.~Rev., D50, 2662, 1994,
hep-th/9312172; D.~Garfinkle, S.B.~Giddings and A.~Strominger,
\journal Phys.~Rev., D49, 958, 1994.}
\lref\banrev{T.~Banks, to appear.}
\lref\cash{Y.~Aharanov, A.~Casher, and S.~Nussinov, \journal
Phys.~Lett., B191, 51, 1987; R.D.~Carlitz and R.S.~Willey, \journal
Phys.~Rev., D36, 2336, 1987.}
\lref\suss{L.~Susskind, {\sl Comment on a Proposal by Strominger},
Preprint SU-ITP-94-14, hep-th/9405103.}
\lref\ande{ B. DeWitt and A. Anderson, \journal Found. Phys., 16, 91, 1986.}
\lref\worm{ S.~Coleman, \journal Nucl.~Phys., B307, 864, 1988; S.B.~Giddings
and A.~Strominger, \journal Nucl.~Phys., B307, 854, 1988.}
\lref\wrloss{A.~Strominger, \journal Phys.~Rev.~Lett., 52, 1733, 1984;
S.W.~Hawking, \journal Phys. ~Lett., B195, 337, 1987.}
\lref\hawktwo{S.W.~Hawking, \journal Phys. Rev., D14, 2460, 1976.}
\lref\bc{T.~D.~Chung and H.~Verlinde, \journal Nucl.~Phys., B418, 305, 1994,
hep-th/931107, S. Das and S. Mukherji, \journal
Phys.~Rev., D50, 930, 1994; A.~Strominger and L.~Thorlacius,\journal
Phys.~Rev., D50, 5177, 1994,
hep-th/9405084.}
\lref\rst{J.G.~Russo, L.~Susskind, and L.~Thorlacius, \journal
Phys.~Rev., D46, 3444, 1993; \journal Phys.~Rev., D47, 533, 1993.}

\Title{\vbox{\baselineskip12pt\hbox{UCSBTH-94-34}\hbox{hep-th/9410187}
}}
{\vbox{\centerline { Unitary Rules for Black Hole Evaporation}
}}
\centerline{{\ticp Andrew Strominger }\footnote{$^\dagger$}
{Email address:
andy@denali.physics.ucsb.edu}
}
\vskip.1in
\centerline{\sl Department of Physics}
\centerline{\sl University of California}
\centerline{\sl Santa Barbara, CA 93106-9530}

\bigskip
\centerline{\bf Abstract}

Hawking has proposed non-unitary rules for computing the
probabilistic outcome of black hole formation.  It is shown that the
usual interpretation of these rules violates the superposition
principle and energy conservation.  Refinements of Hawking's rules are
found which restore both the superposition principle and energy
conservation, but leave completely unaltered Hawking's prediction of a
thermal emission spectrum prior to the endpoint of black hole
evaporation.  These new rules violate clustering.  They
further imply the existence of superselection
sectors, within each of which clustering is restored and
a unitary $S$-matrix is shown to exist.

\Date{}
\newsec{Introduction: Low-Energy Approach to Black Hole Formation/Evaporation}
Consider the formation of a black hole of mass $M$, where $M$ is much
greater than the Planck mass $M_p$, from the collapse of low-energy
({\it i.e.} sub-planckian) matter.  Everyone agrees that the
black hole will evaporate due to Hawking radiation\hawk, and that virtually
all (up to corrections of order $\frac{M_p}{M}$) of the outgoing
quanta will have energies well below $M_p$.  Since both the in- and
out-states can be adequately described by a low-energy, sub-planckian,
effective field theory, it is natural to seek a low-energy effective
description of the scattering interaction. However, there is no
guarantee that this effective description can be derived from the laws
of low-energy physics alone, since the dynamics of gravitational
collapse inexorably lead to regions of high curvature where Planck-scale
physics is important.  Nevertheless, it should be possible to summarize
our ignorance about Planck scale physics in a phenomenological boundary
condition (or generalization thereof) which governs how low energy
quanta enter or exit the planckian region.

In principle this effective description should be derived by a
coarse-graining procedure from a complete theory of quantum gravity such
as string theory.  But this is not feasible in practice.  Instead we
shall find that
the possible
descriptions can be highly constrained by low-energy considerations alone.
Using this latter approach, we shall be led to a new and
satisfying effective description of black hole
formation/evaporation\polst.

A classic example of this type of approach is the analysis of the
Callan-Rubakov effect
\refs{\caru,\jpia}, in which charged $S$-wave fermions are scattered off of a
$GUT$ magnetic monopole.  Even at energies well below the $GUT$ scale, the
scattering cannot be directly computed from a low energy effective field
theory, because the fermions are inexorably compressed into a small region in
the monopole core in which $GUT$ interactions become important.
Initially the $GUT$ scale physics was analyzed in some detail.
The results were then coarse-grained and summarized in an effective boundary
condition for fermion scattering at the origin.  It was subsequently realized
that the detailed $GUT$ scale analysis was largely unnecessary for
understanding the low-energy scattering:
up to a few free parameters (a matrix in flavor space) the effective
description is determined by low-energy symmetries.
We now turn to the
black hole problem with this philosophy in mind.

Classically, black holes are stable, but quantum mechanically they
slowly evaporate and shrink \refs{\hawk}. Hawking has calculated the
outgoing radiation state using low-energy effective field theory
together with the adiabatic
approximation. Although they have certainly been
questioned\foot{
The author's view on this controversy can be found in \aslh.}, both of these
approximations would seem to be
valid as long as the black hole mass $M$  is well above the Planck mass
$M_p$.  The
calculation requires (by causality) only the exterior black hole
geometry.  It follows \hawktwo\ that the outgoing Hawking radiation carries
little
information about what has fallen into the black hole, at least prior to
the evaporation ``endpoint'' at which the black hole shrinks to the Planck mass
and the
approximations break down.  For example, in a theory with an exact
chocolate-vanilla
flavor symmetry, the outgoing radiation prior to the endpoint is {\it
identical}
for black holes formed from vanilla or chocolate matter, and so
information about the flavor of the initial state cannot be obtained
from this radiation.

Once the black hole reaches the Planck mass, quantum gravity must be
solved to continue the evolution.  As quantum gravity is poorly
understood, it might seem that one should simply give up on the problem
at this point.  However, as discussed above,
it still makes sense to ask what a low-energy
experimentalist who makes black holes and measures the outgoing
radiation could observe, and to try to describe this by an effective
field theory. In the following  sections we discuss several possibilities.

\newsec{Remnants?}
One logically possible outcome of gravitational collapse is that planckian
physics shuts off the Hawking
radiation when the black hole reaches the Planck mass, and the
information about the initial state is eternally stored in a planckian
remnant.  As there are infinite numbers of ways of forming black holes
and letting them evaporate, this remnant must have an infinite number of
quantum states in order to encode the information in the initial state.
In an effective field theory
these remnants would resemble an infinite number of species of
stable particles.

This raises the so-called ``pair-production problem''. Since the
remnants carry mass\foot{Massless remnants would create even worse
difficulties.}, it must be possible to pair-produce them in a
gravitational field.  Naively the total pair-production rate
is proportional to the number of remnant species, and therefore infinite.
It is easy to hide a Planck-mass particle, but it is hard to hide an
infinite number of them. Thus it would seem that remnants can be
experimentally ruled out by the observed absence of copious pair-production.

The error in this logic was pointed out in \refs{\bos}, which
may be briefly summarized as follows\foot{In \bos\ dilaton rather than
Reissner-Nordstrom black holes were considered. As
stressed in \refs{\stas}, the argument is cleaner in the
Reissner-Nordstrom context because regions of strongly coupled or
planckian dynamics can be avoided.}. In \stas\ it was shown that
the quantum versions of the charged Reissner-Nordstrom solution
have an infinite degeneracy of stable ``remnant states'' which
for large charge can (unlike their neutral planckian cousins discussed above)
be described with weakly-coupled, semiclassical perturbation theory.
This infinite degeneracy potentially leads to unacceptable
pair-production, so the  Reissner-Nordstrom remnants
provide a good laboratory for analyzing the pair-production
problem\foot{This was also stressed in \sbg.}.
It can be seen that the  different
remnant
states differ solely
by the action of a local operator in a region which is near
or inside of a horizon and causally disparate from
the external observer. Causality ({\it i.e.} the fact that operators
commute at spacelike separations) implies that the causally disparate
states can not be distinguished in a finite-time scattering experiment.
This is certainly at odds with the  naive low-energy
effective description in which each remnant state is
represented as a distinct particle species, and could therefore
all be distinguished {\it e.g.} in finite-time interference experiments.
The naive description
must therefore
fail\foot{A more subtle type of low-energy
effective description, such as in \dxbh\ where the remnant
interior is described by an entire two-dimensional field theory, may still
work.}. This failure can be traced to ultra non-local
(in time) interactions along the remnant worldline. One may
also expect that the infinite degeneracy of states lying in a
causally distant region could not have a divergent effect on any
finite-time pair production process.
This expectation was borne out in
the euclidean instanton calculation of the pair-production rate
in \refs{\garst}, which
yielded a finite result. While certainly
more remains to be understood on this
topic, it is clear that the standard argument that infinite
pair-production is inevitable for all types of remnants is too naive, and
it is plausible that in some theories the pair production rate is
finite.  Further discussion can be found in \refs{\sbg,\remn} and the reviews
\refs{\banrev,\aslh}.

A more inescapable objection to eternal remnants is the lack of any
plausible mechanism to stabilize them.  In quantum mechanics what is not
forbidden is compulsory. In the absence of a conservation
law it is hard to understand why matrix elements connecting a massive
remnant to the vacuum plus outgoing radiation should be exactly zero.
Nature contains no example of such unexplained zeroes. Moreover, a
formal representation of quantum gravity as a
sum-over-geometries-and-topologies certainly includes such processes.
Eternal remnants are therefore highly unnatural.

An alternative possibility is that the ``Planck soup'' which forms when
the black hole reaches the Planck mass continues to radiate in a manner
governed by planckian dynamics until all
the mass is dissipated. In principle, as we do not understand the
dynamics, the radiation emitted by the Planck soup could be correlated
with the earlier Hawking emissions and return all the information
back out to infinity. Energy conservation implies that the total energy
of the radiation emitted by the Planck soup is itself of the  order of the
Planck mass, and thus small relative to the initial mass of the black
hole.  It is very hard to encode all the information in the initial
state with this small available energy.  The only way to accomplish this
is to access very low-energy, long-wavelength states, which requires a
long decay time.  This leads to a lower bound of $\tau \sim M^4$ (in
Planck units) for the decay time of the Planck soup \refs{\cash}. For a
macroscopic black hole this far exceeds the lifetime of the universe.
Hence, it is not possible for the information to be emitted in a
planckian burst at the end of the evaporation process.  In this scenario
one necessarily has a long-lived, but not eternal, remnant. Note that our
discussion
required no knowledge of planckian dynamics.  This is a prime example of
how low-energy considerations highly constrain the possible
outcome of gravitational collapse.

Of course, long-lived remnants are implausible without an explanation
for their long lifetime, or a mechanism for the Planck soup to reradiate
the information.  We shall encounter both below.

\newsec{Information Destruction}
\ifig\fone{Collapsing radiation forms an apparent
black hole (shaded
region) which evaporates, shrinks down to $r=0$ at $x_E$, and
subsequently disappears.  The dashed wavy line is the region at which
Planck-scale physics becomes important, and is just prior to the
classical singularity. According to Hawking, information which crosses
the event horizon is irretrievably lost.}
{\epsfysize=4.50in \epsfbox{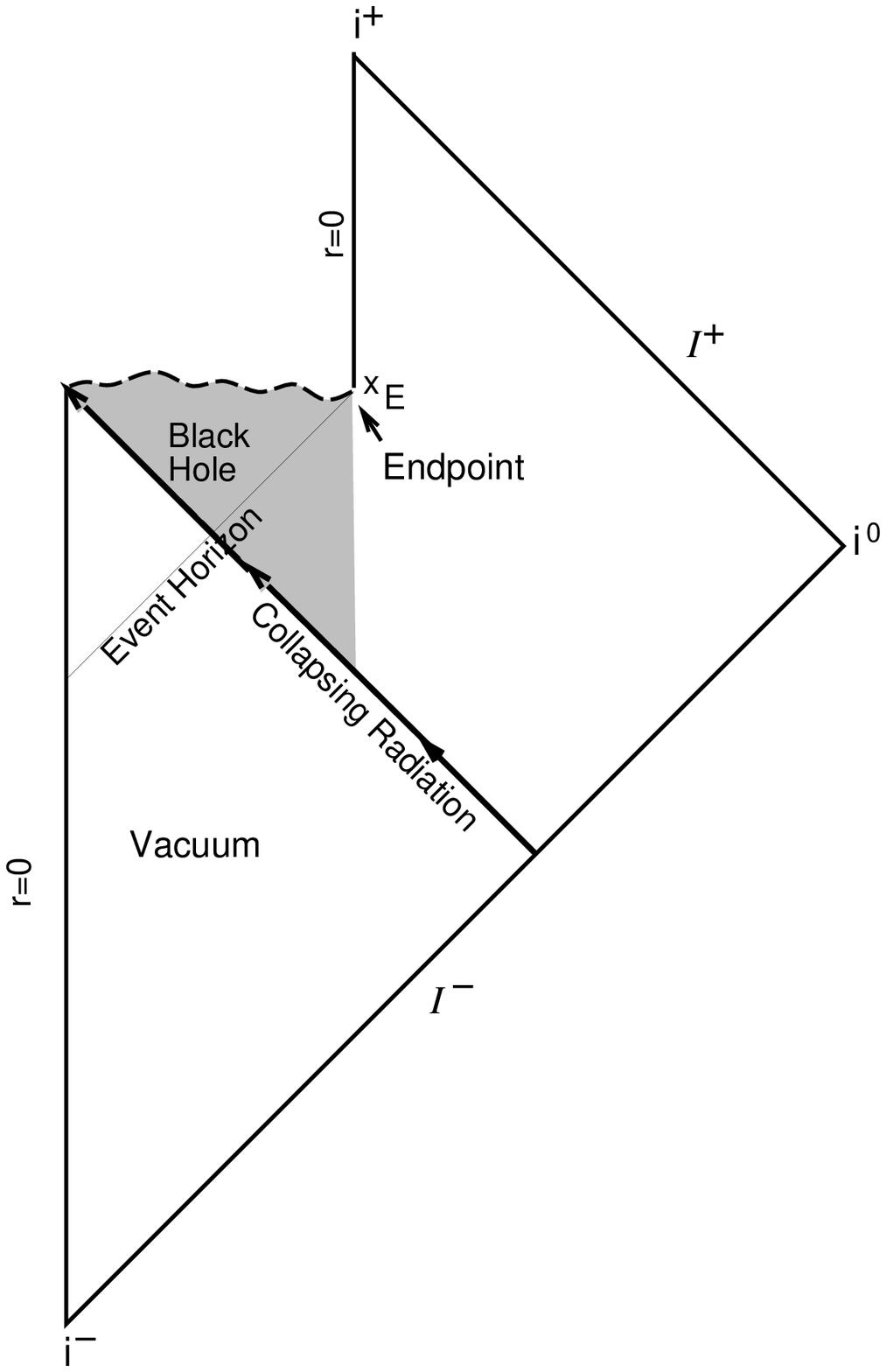}}

Faced with the apparent unpalatability of remnants, Hawking argued in
favor \hawktwo\ of a different possibility, depicted in \fone.  The black hole
disappears in a time of order the Planck time after shrinking to the
Planck mass, and the infalling information disappears with it.  After
all, in practice, information often escapes to inaccessible regions of
spacetime, even in the absence of gravity.  The inclusion of gravity,
Hawking argues, implies information is lost in principle as well as in
practice.

\ifig\ftwo{Hawking's rule for density matrix superscattering
for single
black hole formation.  The left (right) side of the diagram represents
the evolution of the ket(bra) of the density matrix.  The trace over
the
part of the Hilbert space which falls into the black hole is schematically
represented
by sewing together the left and right black hole interiors.}
{\epsfysize=2.50in \epsfbox{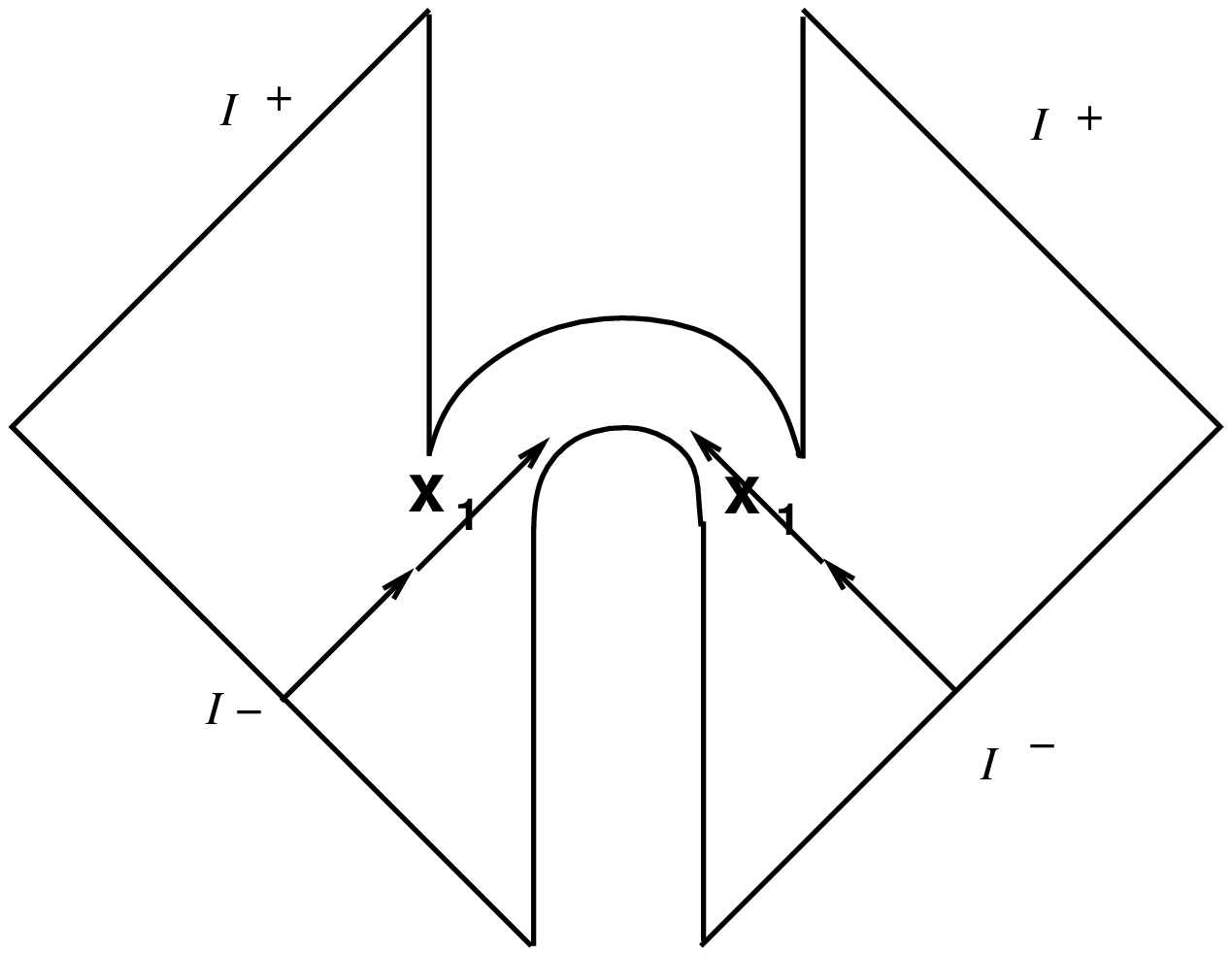}}

Since information is lost in this proposal, there can be no unitary
$S$-matrix mapping in-states to out-states.  Rather, Hawking suggests that a
``superscattering'' matrix, denoted ``$ {\not\kern-0.2em S} $'', which maps
in-density matrices (of the general form $ \rho=\sum \rho_{ij}|\psi_i\rangle
\langle \psi_j|$ )
to out-density matrices can be constructed as
\eqn\dss{\not\kern-0.2em S =tr_{BH} S\, S^\dagger\ .}
$ {\not\kern-0.2em S} $ will not in general preserve the entropy
$-{ tr}\rho \ln \rho$.
In components, $ {\not\kern-0.2em S} $ acts on an in-density matrix
as $\bigl(  {\not\kern-0.2em S}
\bigl[ \rho \bigr] \bigr)_{kl}=\bigl( {\not\kern-0.2em S}
 \bigr)_{kl}^{~~~ij}\rho_{ij}$.
$S$ here is a unitary operator which maps the in-Hilbert space to the
product of the out-Hilbert space with the Hilbert space of states which
falls into the black hole (defined, for example, as quantum states on
the event horizon in \fone ). $tr_{BH}$ is the instruction to trace
over these latter unobservable states.  Expressions of the form \dss
\ are familiar in physics, and arise, for example, in the computation of
$e^+e^-$ scattering in which the spins of the final state are not
measured. A diagrammatic representation of Hawking's prescription for
the case of one black hole appears in \ftwo .

\ifig\fthree{Hawking's rule for superscattering of two black
holes involves two
traces, one for each black hole.}
{\epsfysize=2.50in \epsfbox{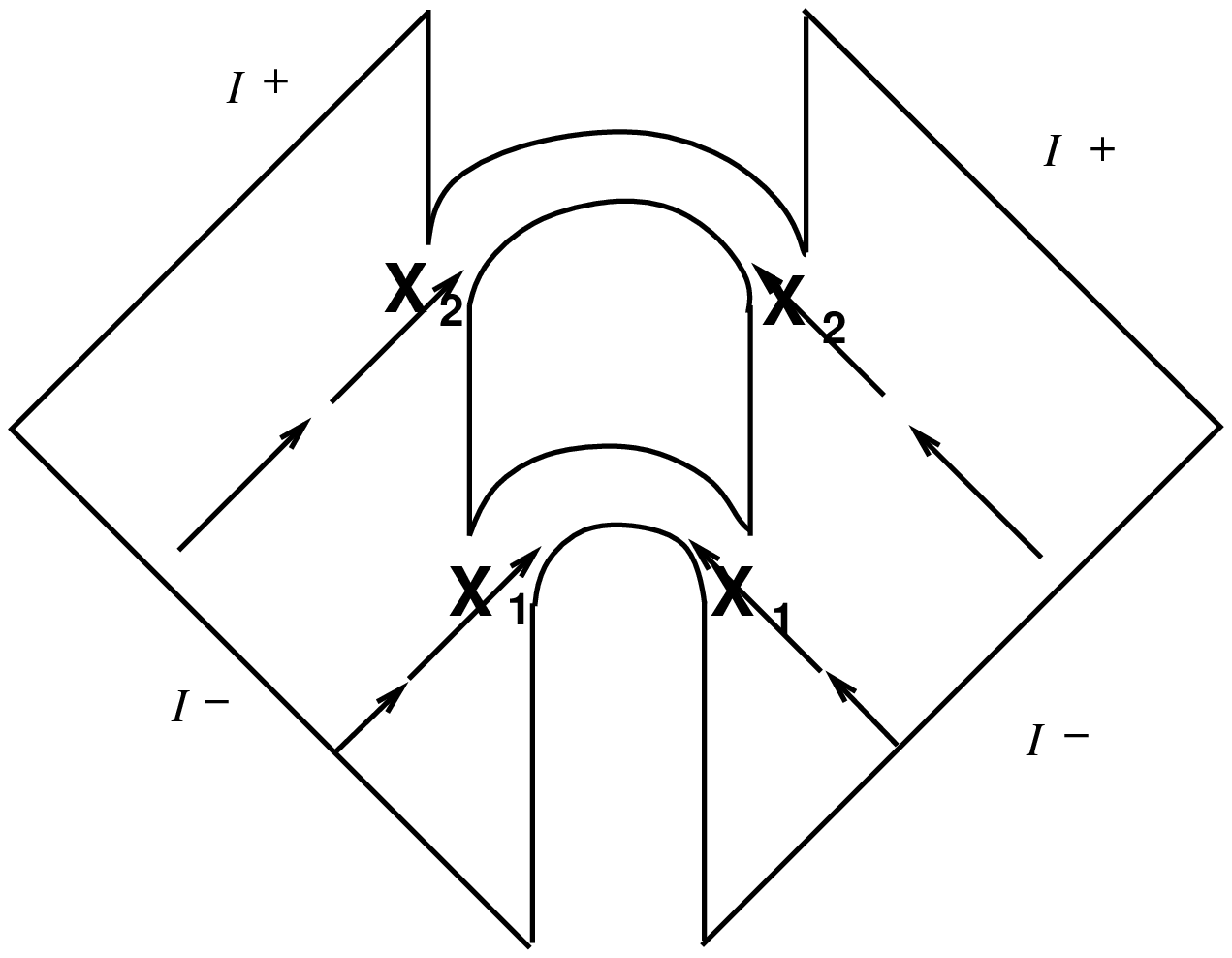}}

It is implicit in Hawking's proposal that the probabilistic outcome of
the formation/evaporation of an isolated black hole near the spacetime location
$x_1$ can in
this manner be computed from the portion of the quantum state which
collapses to form the black hole. In this case
the outcome of forming a second black
hole at a greatly spatially or temporally separated location $x_2$
is uncorrelated and the two-black hole $ {\not\kern-0.2em S} $-matrix can be
decomposed into a product of single black hole $\not\kern-0.2em S$-matrices
 (In other words, probabilities cluster.) The
corresponding diagrammatic representation of $\not\kern-0.2em S$ for the case
of two
black holes is given in \fthree .

\newsec{The Superposition Principle}
In fact as it stands Hawking's
proposal is in conflict with the superposition
principle\foot{The arguments of this and the following section
may be related to those employed
in a somewhat different context in \bps\ and \sbg.}.  To see this
note that there are inevitably
non-zero but possibly small quantum fluctuations in the location $x_1$
where the black hole is formed.  $tr_{BH}$ instructs one to trace by
equating the black hole interior states of the bra and the ket in
the density matrix, independently of the precise location where the
black hole is formed.  Now $x_1$ cannot be an
observable of the black hole interior Hilbert space, since  by
translation invariance the interior
state of the black hole does not depend on where it was formed. Hence
 the trace will
include contributions from black holes interiors which are in the same
quantum state, but which were formed at slightly different spacetime
locations.

\ifig\ffour{
Superscattering of an initial coherent superposition of
semiclassical states which form black holes near widely separated
locations $x_1$ and $x_2$.  The superposition principle and translation
invariance imply that all four diagrams contribute.}
{\epsfysize=3.50in \epsfbox{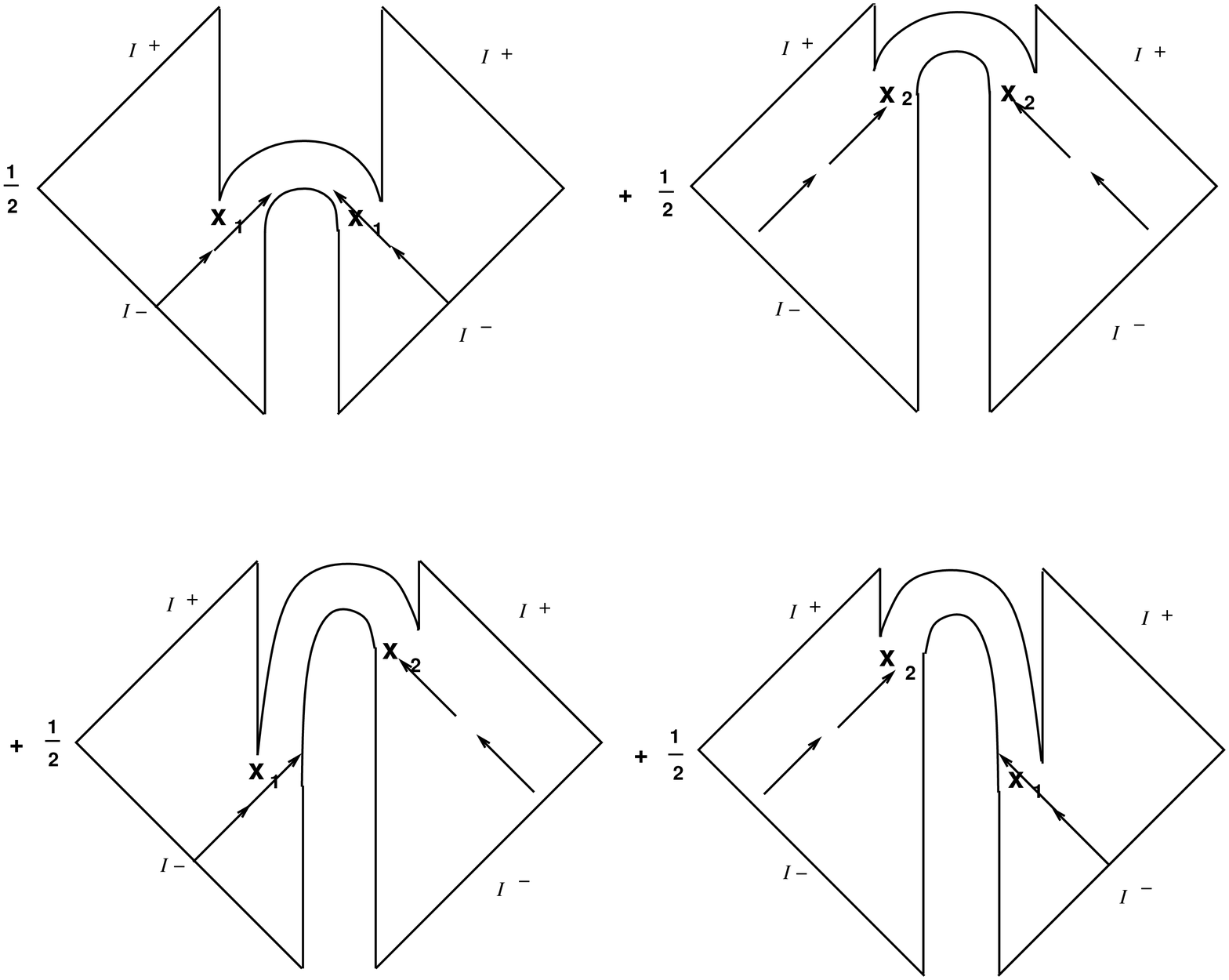}}

This phenomenon is more pronounced in initial states for which the
fluctuations in the location of the black hole are not small. Such states
can certainly be constructed.  For
example, let the in-state be the coherent superposition
\eqn\two{|\psi_{\rm in}\rangle= \frac{1}{\sqrt{2}} \left(|x_1\rangle +
|x_2\rangle\right),}
where $|x_i\rangle$ is a semiclassical initial state which collapses to
form a black hole near $x_i$, and $x_1$ and $x_2$ are very widely
separated spacetime locations.  By continuity
the construction of $\not\kern-0.2em S$
must include terms which equate the interior black hole
bra-state formed at $x_1$ with the ket-state formed at
$x_2$.  There are then four terms in $\not\kern-0.2em S$ as
illustrated in \ffour .

\ifig\ffive{The superposition principle implies that for two
black holes
this cross diagram must be added to that of \fthree , correlating
widely
separated experiments.}
{\epsfysize=2.50in \epsfbox{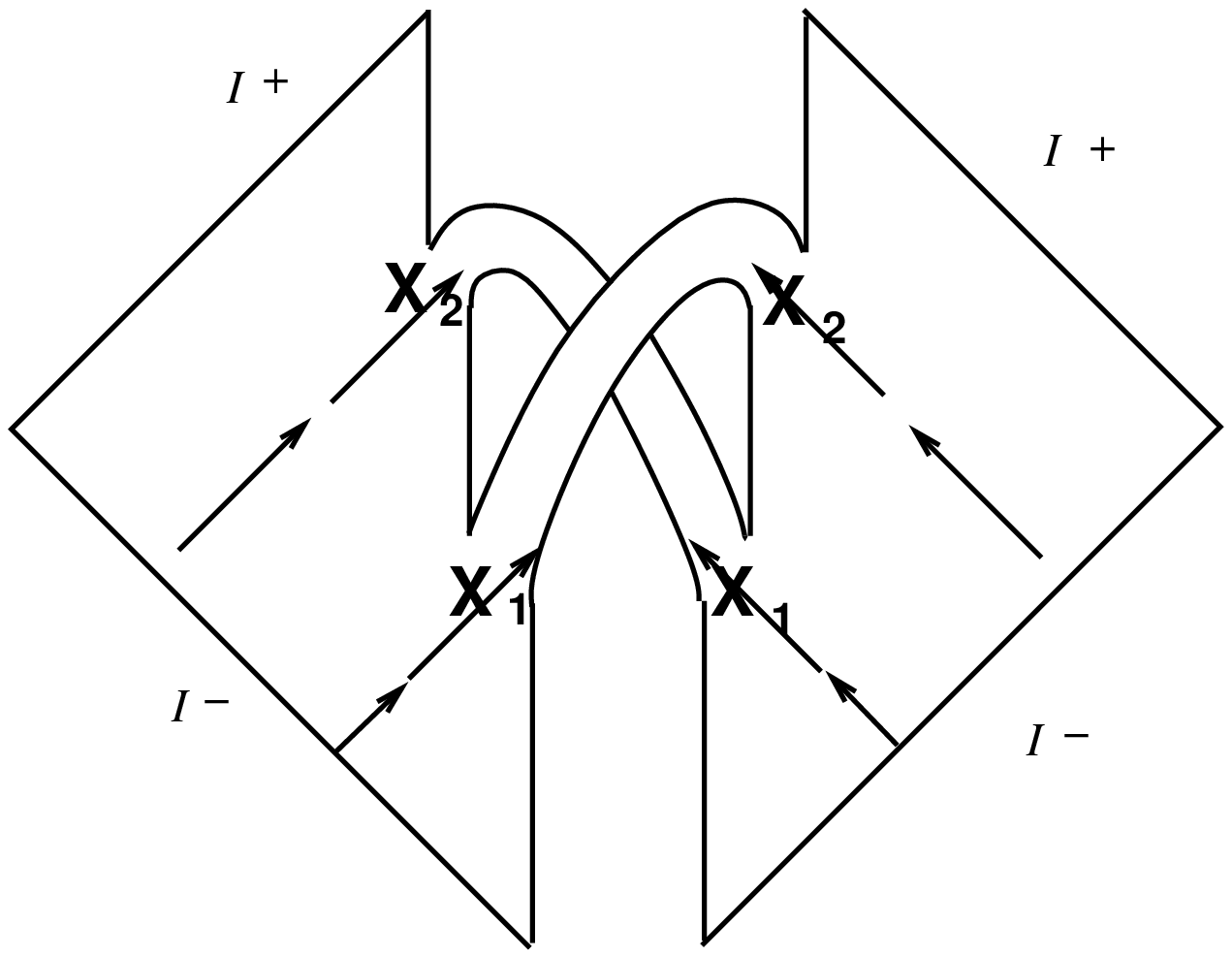}}

It may already seem rather strange
that $\not\kern-0.2em S $  should contain such
correlations between widely separated events, but matters become even
worse when one considers a semiclassical initial state $|x_1, x_2\rangle$
which collapses to form two black holes at the widely separated
locations $x_1$ and $x_2$.  The superposition principle then requires
that the cross diagram of \ffive\ be added to the diagram
of \fthree \foot{This extra  cross diagram will be
small if the parts of the incoming
states which form the two black holes are
very different and the black hole interiors have a correspondingly
small probability of being in the same state. On the other hand if they differ
only by a translation, \ffive\ will be similar in size to \fthree.}.
To see this, consider a smooth
one-parameter family of initial states $|x_1(s), x_2(s)\rangle$ in which
the locations $x_1$ and $x_2$ are interchanged as the parameter $s$ runs
from zero to one.  Let the in-state be
\eqn\three{|\psi_{\rm in}\rangle= \int^1_0 ds|x_1(s),
x_2(s)\rangle\ .}
Then the diagrams of \fthree\  and \ffive\ are interchanged as $s$ goes from
0 to 1 in the ket-state, so neither can be invariantly
excluded.

Thus the superposition principle implies that one cannot, in
the manner Hawking suggests, compute the probabilistic outcome of a
single experiment in which a black hole is formed.  Knowledge of all
past and future black hole formation events is apparently required
to compute the superscattering matrix
(although we shall see below that this is not as unphysical as it seems).
Again, it is striking that low-energy reasoning highly constrains
possible outcomes of black hole formation without requiring knowledge of
 planckian dynamics.

\ifig\fsix{When the evolution of spacelike slices (denoted by the dashed lines)
reaches the endpoint
$x_E$, the incoming slice, and the quantum state on the slice,
is split into exterior and interior portions.
This splitting process is described using the operator $\Phi_J$ ($\Phi_K$)
which annihilates (creates) an incoming (outgoing)
asymptotically flat slice in the $J'th$ ($I'th$) quantum state and $\Phi_i$
which creates an interior slice in the $i'th$ quantum state.}
{\epsfysize=3.50in \epsfbox{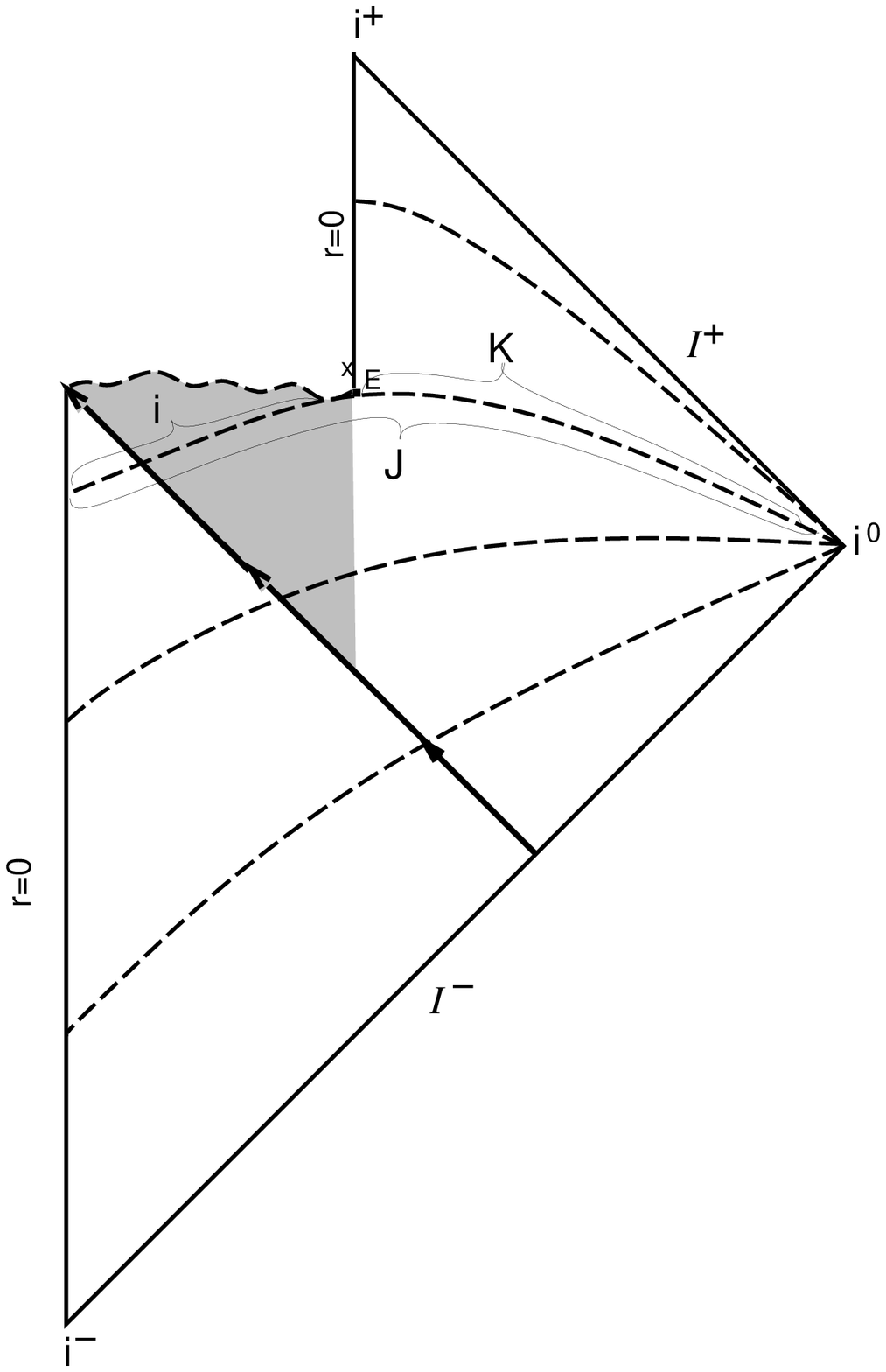}}

Note that our conclusions about difficulties with the usual
interpretation of Hawking's proposal
have derived from consideration of {\it superpositions} of semiclassical
states which form black holes.  These difficulties have not been so
evident in previous discussions simply because such superpositions are
not usually considered.

\newsec{Energy Conservation}
Although the superposition principle is restored with the extra cross diagram
of \ffive , correlations are introduced between arbitrarily widely separated
experiments, and clustering is violated \refs{\suss}.
Thus we seem to be faced with a choice:
abandon the superposition principle or abandon
clustering.  In fact we shall see below
that the breakdown of clustering
is a blessing in disguise, but first we need to
introduce a second refinement of Hawking's prescription required by
energy conservation\foot{I am grateful to S. Giddings for emphasizing
to me the
importance of understanding energy conservation in this context.}.
\ifig\fseven{Anderson and DeWitt studied a free field
propagating on a
geometry which is split into two at time $t=t_s$ by reflecting boundary
conditions at $x=0$. The sudden change in the Hamiltonian produces
infinite energy pulses which propagate along the dashed lines.}
{\epsfysize=3.00in \epsfbox{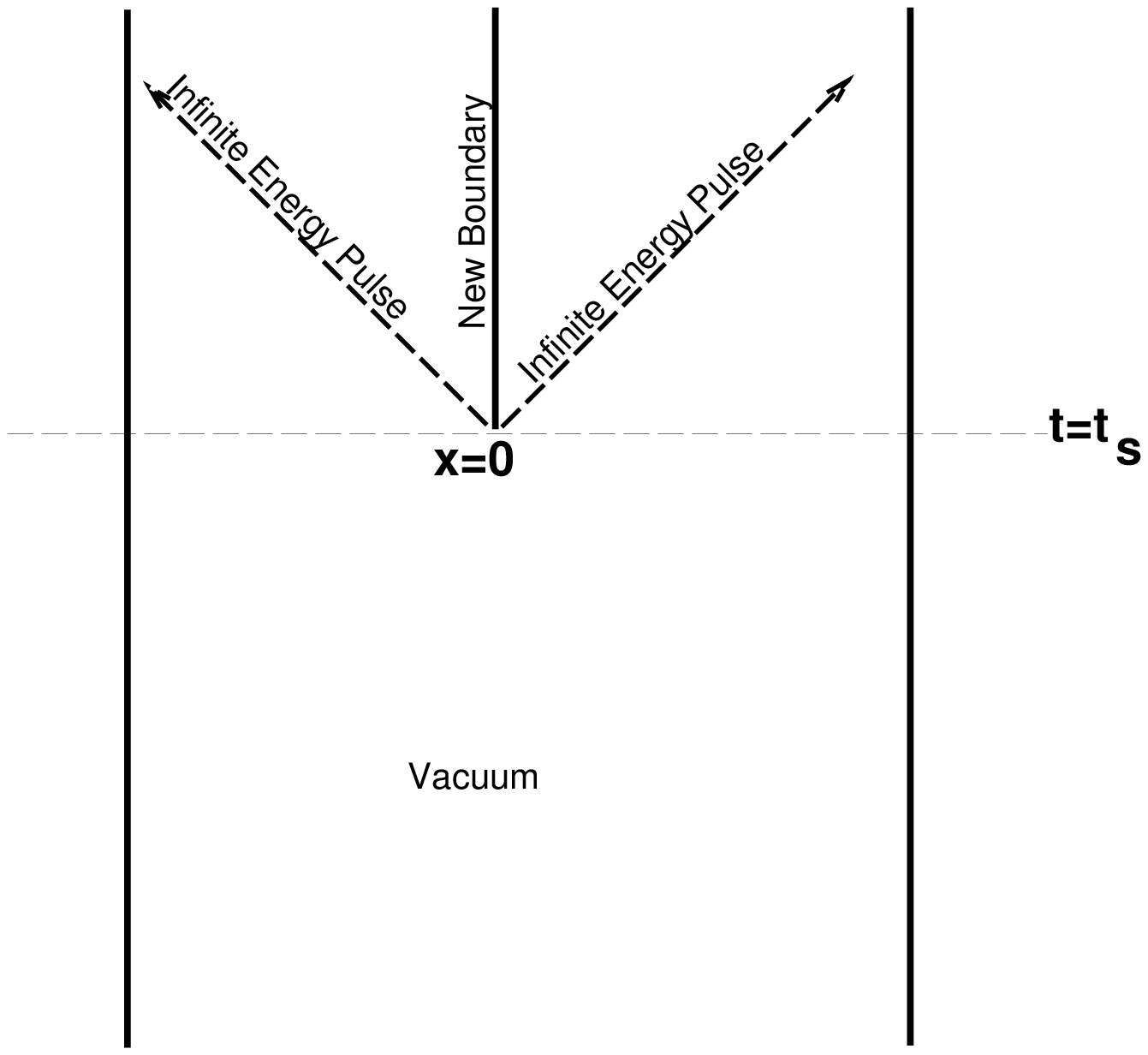}}
In computing the $\not\kern-0.2em S$-matrix, complete
spacelike slices are split into interior and exterior portions when they
encounter the evaporation endpoint at $x_E$, as illustrated in \fsix .
One imagines that the Hilbert
space on these slices is also split into the product of two
corresponding interior and exterior Hilbert spaces.  This requires some
new boundary conditions originating at $x_E$: an incoming light ray just prior
to $x_E$
falls into the black hole, while an incoming light ray just after $x_E$
reflects through the origin and back out to null infinity. (Explicit examples
of
such boundary conditions exist in $1+1$ dimensions \refs{\bc}, but for our
present purposes an explicit form will not be needed.) Implementing
this in practice immediately runs afoul of the Anderson-DeWitt
\refs{\ande} problem.  These authors considered the propagation of a free
conformal field in $1+1$ dimensions on the trousers spacetime of \fseven\
in which (as in the black hole case) spacelike slices are split into
two portions at some fixed time $t_s$, when reflecting boundary
conditions are turned on at $x=0$. They find that the vacuum state for $t<t_s$
evolves to a state with infinite energy for $t>t_s$.  This is not
surprising since the Hamiltonian changes at an infinite rate at $t=t_s$.

This phenomenon is not peculiar to two dimensions. A change in the Hamiltonian
in the form of new boundary conditions at a fixed spacetime location
violates general covariance and therefore energy conservation.
This problem should be expected
to affect the separation of Hilbert space into interior and exterior
portions at the evaporation endpoint $x_E$ for the black hole case.
Indeed the most concrete description given of this splitting
process --- that in
the $1+1$ dimensional RST model \rst --- suffers from exactly this problem.
Energy is not conserved in this model because the quantum state of the
matter field acquires infinite energy as it is propagated past $x_E$ \lpst.

\ifig\feight{A cosmic string decays into two pieces which end
at monopoles.  This process conserves energy, and the decay
Hamiltonian involves the fields $\phi_J$ which annihilates the incoming
string and $\phi_I, \phi_K$ which create the two outgoing strings.}
{\epsfysize=3.00in \epsfbox{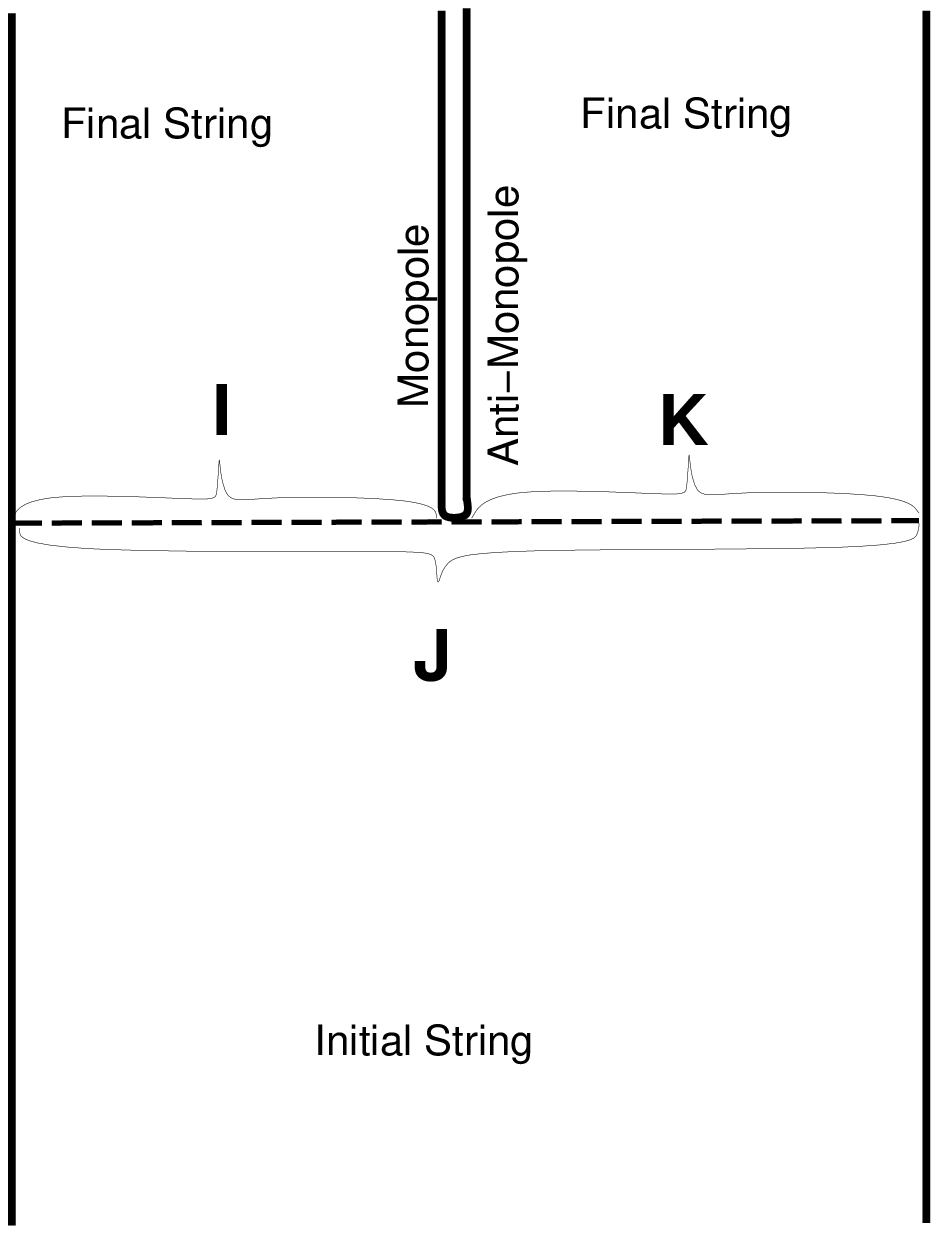}}

To remedy this, a smooth energy-conserving method
of splitting the incoming Hilbert space
into two portions is needed.  A physical example of a system which exhibits
such
a smooth splitting is given by cosmic string decay.  Consider, {\it
e.g.} a magnetic flux tube described by a
Nielsen-Olesen vortex.  At low energies it is
described by a $1+1$ dimensional quantum field theory whose massless
fields are the transverse excitations $X(\sigma)$ of the string.  Next suppose
that
the string can decay by the formation of a heavy monopole-anti-monopole
pair which divides the string into two parts.  Clearly such a process can
occur and will conserve energy.  It cannot, however, be simply described by
propagating the $1+1$ dimensional fields on the fixed geometry of \fseven\
(or superpositions thereof), as analyzed by Anderson and DeWitt.
Rather, the decay rate depends on the
final state after the split through initial and final wave function
overlaps appearing in decay matrix elements, and the decay time is thus
correlated with the quantum states on the two final strings.  This
decay process may be conveniently and approximately (at low energies)
described by the interaction Hamiltonian (see \feight )
\eqn\seven{{\cal H}_{\rm int} = \sum_{I,J,K} g\, \rho_{IJK}\phi_I
\phi_J\phi_K\ .}
In an appropriate basis, the mode of the field operator
\eqn\eight{\phi_I = a_I  + a^\dagger_I}
here creates or annihilates (from nothing) an entire string in the
$I$'th quantum state with wave function $u_I[X(\sigma)]$, and $[a_{I},
a^\dagger_J] = \delta_{I J}$. We emphasize that
$\phi_I$ is {\it not} an operator which acts on the single-string Hilbert
space. $\rho_{IJK}$
is the
overlap of the one initial and two final state wave functions $u_I, u_J, u_K$
for strings aligned as in
\feight.
$g$ is an effective low-energy coupling constant governing the decay
rate, in which our ignorance of the microscopic details of the splitting
interaction is hidden.

Despite many efforts, no other method of avoiding the Anderson-DeWit
problem is known.  We accordingly {\it presume} that the disappearance of a
black hole is properly viewed as a  quantum decay process in which the
black hole interior and exterior are separated. We cannot {\it derive}
this presumption without solving quantum gravity. Nevertheless, it appears
to be forced on us by low energy considerations. We know of no other
consistent effective description.

While it is probably too much to derive this picture from a microscopic
theory of quantum gravity, it might be possible to find a toy
model -- {\it e.g. } in two dimensions or minisuperspace -- in which
it can be consistently realized. Such a model would certainly
be of great interest.

In this picture the decay does not then occur
instantaneously when the semiclassical evaporation endpoint $x_E$ is reached.
Rather the geometry itself decides when to split (some time after $x_E$)
in a quantum mechanical
fashion, controlled by the effective decay coupling constant as well as
phase space factors appearing in initial/final wave function overlaps.
The precise splitting time, like all other quantities, is then
subject to quantum fluctuations and correlated with the final state.
\newsec{The New Rules}
We have proposed two modifications of Hawking's prescription: the
inclusion of cross diagrams as in \ffive\
and the description of the final stages of
black hole evaporation as a quantum decay.  Both of these modifications are
encoded in
the formulae
\eqn\psiev{i\partial_T |\psi(T)\rangle = (H_0 + H_{\rm int}\rangle
|\psi(T)\rangle}
\eqn\drule{\eqalign{\not\kern-0.2em S \bigl[|\psi_{\rm in}\rangle&
\langle \psi_{\rm in} |\bigr] = tr_{BH} |\psi_{\rm out}\rangle
\langle\psi_{\rm out}|\cr
|\psi_{\rm in}\rangle&\equiv |\psi(-\infty)\rangle\cr
|\psi_{\rm out}\rangle&\equiv |\psi(+\infty)\rangle\cr}}
where  $H_0$ is the usual  gravitational Hamiltonian
which evolves the system along a set of spacelike slices labeled by time
coordinate $T$, but does not include the decay interaction.  The latter
is given, in precise analogy to the cosmic string case by
\eqn\eleven{H_{\rm int} = \sum_{i,J,K} g\, \rho_{iJK} \Phi_i \Phi_J \Phi_K\ .}
$\Phi_J$ here creates or annihilates an asymptotically flat spacetime in
the $J$'th quantum state. (It does {\it not} act on the flat space vacuum
to create the $J$'th excitation.) $\Phi_i$ creates or annihilates a
compact spacetime, {\it i.e.} a black hole interior in the $i$'th quantum
state.
$\rho_{iJK}$ is the wave function overlap computed by aligning the
geometries as depicted in \fsix . $g$ is a decay coupling constant in
which our ignorance of Planck-scale physics is hidden.

The operators $\Phi_i=a_i  + a^\dagger_i$ generate a
multi-black-hole-interior Hilbert space $H_{\rm BH}$. If $[a_i,
a^\dagger_j] = \delta_{ij},\ |\psi_{\rm in}\rangle$ is taken to obey
$a_i|\psi_{\rm in}\rangle=0$ and $tr_{\rm BH}$ is the trace over $H_{\rm
BH}$, then the rule \drule\ for construction of ${\not\kern-0.2em S}$  contains
(with the
correct weighting) the cross diagrams required by the superposition
principle.

The $\Phi_i$'s may be simply viewed as a convenient mnemonic
for constructing the diagramatic expansion of $\not\kern-0.2em S$.
 Alternately,
one may think of the black hole interiors as forming baby universes which
inhabit a ``third quantized'' Hilbert space on which the $\Phi_i$'s act.
However, the detailed dynamics of these baby universes will not be
needed for our purposes because we view them as unobservable.
\newsec{Superselection Sectors, $\alpha$-parameters, and the Restoration of
Unitarity}
Next let us suppose that the initial state is in an ``$\alpha$-state''
obeying \worm
\eqn\twelve{\Phi_i | \{\alpha\}\rangle = \alpha_i | \{\alpha\}\rangle}
where the $\alpha_i$'s are $c$-number eigenvalues, rather than $a_i
|\psi_{\rm in} \rangle=0$.
In such a state the operator $\Phi_i$ may be everywhere replaced by its
eigenvalue and
\eqn\thirteen{H_{\rm int} = \sum_{J,K} g_{JK} \Phi_J\Phi_K}
with
\eqn\fourteen{g_{\rm JK} = \sum_{i}\alpha_i\rho_{iJK} = c-{\rm numbers}\ .}
$H_{\rm int}$ {\it reduces to an operator on the Hilbert space of a single
asymptotically flat spacetime.}  It then follows immediately from
\psiev\ that the out-state
\eqn\fifteen{|\psi_{\rm out}\rangle =
S_{\{\alpha\}}|\psi_{\rm in}\rangle}
is a unitary, $\alpha$-dependent transformation $S_{\{\alpha\}}$ of
the in-state. $S_{\{\alpha\}}$ here is obtained by solving \psiev,
which reduces to an ordinary Schroedinger-Wheeler-DeWitt equation in
an $\alpha$-state.

The reader may suppose that this result is of little interest
because the generic state is not an $\alpha$-state, rather it is a
coherent superposition of $\alpha$-states.  To understand the properties
of such superpositions, consider
\eqn\psdc{|\psi\rangle = \theta | \{\alpha\}\rangle + \theta^\prime |
\{\alpha^\prime\}
\rangle}
where
\eqn\aort{\langle\{\alpha\} | \{\alpha^\prime\}\rangle=0}
since $\alpha$-states are eigenstates of a hermitian operation with distinct
eigenvalues.

Observables ${\cal O}_i$ corresponding to measurements in the
asymptotically flat spacetime do not act on the multi-black-hole-interior
Hilbert space $H_{\rm BH}$.  Hence they commute with the $\Phi_i$'s and
leave the
$\alpha$-eigenvalues unchanged.  It then follows from \aort\ that
\eqn\ssct{\langle\{\alpha\}| {\cal O}_i | \{\alpha^\prime\} \rangle = 0}
and
\eqn\ssrt{\eqalign{&\langle\psi | {\cal O}_1 {\cal O}_2 \cdots {\cal O}_N |
\psi\rangle~~~~~~\cr
&~~~~~~ = |\theta |^2 \langle\{\alpha\} | {\cal O}_1 {\cal O}_2 \cdots {\cal
O}_N |\{\alpha\}\rangle\cr
&~~~~~~+|\theta^\prime|^2 \langle\{\alpha^\prime\}| {\cal O}_1 {\cal O}_2
\cdots {\cal O}_N |\{\alpha^\prime\}\rangle\ .\cr}}
A similar relation holds for more general superpositions of
$\alpha$-states, including the ``vacuum'' state obeying
$a_i|\psi\rangle=0$.

The content of \ssrt\ is that the $\alpha$'s label non-communicating
{\it superselection sectors}.  According to \ssrt, the amplitude for
repeating an experiment which measures an $\alpha$-value and obtaining a
different result the second time is
zero.\foot{In the Copenhagen interpretation, one would say that
measurement of an $\alpha$-value collapses the wave function to the
corresponding $\alpha$-eigenstate.} Once an experiment records a given
$\alpha$-value, all future experiments will agree.  There may be
parallel worlds with different $\alpha$-values, but we can never know
about them.  Hence {\it the $\alpha$'s are effectively  constants}
and {\it black hole formation/evaporation is an effectively unitary
process}.\foot{This argument parallels those in earlier work on baby universes.
In \refs{\wrloss}  it was argued, following \hawktwo, that virtual,
planckian baby universes destroy
information.  This conclusion was shown in \refs{\worm} to be false after
proper accounting of superselection sectors.  Following these
developments, many authors tried and failed to adapt the mechanism of
\worm\ to
avoid information destruction by black holes.  The missing ingredient in
these previous attempts to adapt the results of \refs{\worm}
 was the description of the Hilbert space split
as a quantum mechanical decay process.}

We find this result extremely satisfying.  Having modified Hawking's
superscattering rules so as to comply with the superposition principle and
energy conservation, we see that unitary is restored as a free bonus.
This attests to the robust nature of quantum mechanics, and the inherent
difficulty in finding self-consistent modifications.

The real significance of the very-long-range correlation produced by the
cross diagram of \ffive\ is now evident.  They simply conspire to
produce
infinite-range correlations between $\alpha$-values measured in widely
separated experiments.  They do {\it not} allow messages to be sent faster
than the speed of light, or money to be consistently won at the racetrack.

What are the $\alpha$'s in our universe?  Even an exact solution to
string theory could not answer this question: They can only be
determined by forming black holes and measuring the out-state.  Until
they are known, the outcome of gravitational collapse is unpredictable.
The time reverse of this statement is that
information is lost in the sense that the in-state which formed a black
hole cannot be determined even from complete knowledge of the out-state.
This is certainly similar to, and could be regarded as a refinement of,
Hawking's original contention that information is lost in black hole
processes.
Indeed, if one performs a Gaussian average over $\alpha$'s one
recovers results similar to Hawking's (in that pure states
go into mixed ones)
for the case of a single black hole.  Thus the
difference between our proposal and Hawking's is in practice quite
subtle.

The following analogy may clarify the situation.  Consider scattering
photons off of a hydrogen atom.  Imagine that QED is perfectly
understood, except that the value of the fine structure constant is
unknown.  In this case it will not be possible to predict (retrodict)
the out-state (in-state) from the in-state (out-state) of a single
experiment, so that in a sense one could say that information is lost.
However, after performing many scattering experiments, the fine
structure constant is effectively measured, and no further information
loss occurs.

Information loss in black hole formation/evaporation is of exactly this
type.  It does not arise from a fundamental breakdown of unitarity,
rather it is associated with a lack of knowledge of coupling constants
(the $\alpha$'s or $g_{JK}$'s).  The only difference is that in the QED
case there was only one relevant coupling, while in the black hole case
many are needed (more than $e^{4\pi M^2}$ \polst\ ) even to predict the outcome
of
a single fixed in-state, and an enormous number of experiments would be
required to actually measure the parameters. Indeed, since there are an
infinite number of in-states which form black holes (of unrestricted
mass), it is never possible  to measure {\it all} the $\alpha$
parameters.

The alert reader may be concerned about the status of the
information/energy bounds discussed earlier, which constrain the rate at
which the information can be returned with the
small amount of energy available near and after the endpoint. The
arguments for these bounds are quite general and certainly apply to our
proposal.  Thus unitarity implies that our decay rate must be very slow.
One cannot simply
 explain this with a small $g$ as $g$ --- though hard to calculate ---
 is naturally order one
in Planck units. Rather it was shown explicitly in a two dimensional model
in \refs{\polst} that
the decay is highly
suppressed by phase space factors: due to entanglement of the interior
and exterior states, the overlap between the initial and final state
wave function is small, providing for compatibility with the
information/energy bounds (see also \sbg ). Unitarity implies a similar phase
space suppression
in four dimensions, but this remains to be analyzed in detail.
\newsec{Conclusion}
In conclusion, Hawking's proposal for information destruction by black
holes --- as usually interpreted ---  violates
in addition to unitarity, the superposition principle
and energy conservation.  Refinements of (or reinterpretations of)
his proposal which restore the superposition principle and energy
conservation automatically restore unitarity, after the existence of
superselection sectors is properly accounted for.  This can be
accomplished without requiring that planckian dynamics become important
at low curvatures (as some have advocated): The
description agrees exactly with Hawking's everywhere that
semiclassical
reasoning is valid, namely prior to the evaporation endpoint.
It also does not invoke the existence of stable objects with no natural
right to eternal life: Rather it predicts the existence of long-lived
remnants whose long
lifetime may be naturally explained by phase-space suppression of the decay
rate. Thus a unitary, causal
description of black hole formation/evaporation appears to be
compatible
with all known constraints of low-energy physics.

\centerline{\bf Acknowledgments}

This is an expanded version of a talk given at the Seventh Marcel
Grossman Meeting on General Relativity in July 1994, and will appear
in the proceedings.  The first half
presents new results (not
published elsewhere) while the second half describes recent work with
J.~Polchinski \refs{\polst}.
This work was supported in part by DOE grant DOE-91ER40618.  I am
grateful to A. Anderson, T.~Banks, K. Becker, M. Becker,
S. Coleman, S.~Giddings, S.~Hawking, D.~Lowe, J.~Polchinski,
J.~Preskill, M.~Srednicki, L.~Susskind, L.~Thorlacius and V. Rubakov
for useful
discussions.

\listrefs
\end